# A Lightweight Inception Boosted U-Net Neural Network for Routability Prediction

Hailiang Li, Yan Huo, Yan Wang, Xu Yang, Miaohui Hao, Xiao Wang

*Integrated Circuit Enabling Technology (ICET) group of Advanced Electronic Components and Systems (AECS) Division*
*Hong Kong Applied Science and Technology Research Institute Company Limited* (ASTRI)
Hong Kong, China
Email: {harleyli, jennyhuo, yanwang, xuyang, miaohuihao, ericwang}@astri.org

*Abstract*—As the modern CPU, GPU, and NPU chip's design complexity and transistor counts keep increasing, and with the relentless shrinking of semiconductor technology node to nearly 1 nanometer, the placement and routing have gradually become the two most pivotal processes in modern very-large-scale-integrated (VLSI) circuit back-end design. How to evaluate routability efficiently and accurately in advance (at the placement and global routing stages) has grown into a crucial research area in the field of artificial intelligence (AI) assisted electronic design automation (EDA). In this paper, we propose a novel U-Net variant model boosted by an Inception embedded module to predict Routing Congestion (RC) and Design Rule Checking (DRC) hotspots. Experimental results on the recently published CircuitNet dataset benchmark show that our proposed method achieves up to 5% (RC) and 20% (DRC) rate reduction in terms of Avg-NRMSE (Average Normalized Root Mean Square Error) compared to the classic architecture. Furthermore, our approach consistently outperforms the prior model on the SSIM (Structural Similarity Index Measure) metric.

## I. INTRODUCTION

Artificial intelligence (AI) is driving innovation in the development of central processing units (CPUs), graphics processing units (GPUs), and neural processing units (NPUs). The emerging deep learning (DL) technologies have brought new vitality to computer-aided design (CAD) applications for very-large-scale-integrated (VLSI) circuit design. These advances also enable faster design [4, 5] and more accurate cross-stage prediction [2, 3], significantly improving the design closure and the final quality of results (QoR). The modern VLSI design process encompasses the front-end design and back-end design stages. In addition, the typical back-end design includes floorplanning, placement, clock tree synthesis, global routing, detailed routing, timing closure, etc., which maps a synthesized netlist (derived from the front-end design) to a physical GDS (Graphic Design System) for the final tape-out. Statistics reveal that the placement and routing phases are the top two time-consuming steps in the EDA back-end design flow. Meanwhile, routing congestion can significantly affect Performance, Power, and Area (PPA) metrics. And design rule violations may directly lead to unmanufacturable designs. Therefore, it is meaningful and critical to accurately and efficiently evaluate the placement results before routing. As a result, the routability prediction has become one of the key research areas in modern VLSI circuit design in the past decade. However, existing architectures [2, 3] for routability prediction tasks still need to be improved in terms of accuracy, while recently the revised models [15, 16, 17] are running slow in computationally demanding scenarios.

In this paper, we propose a novel lightweight Inception [10] boosted U-Net architecture neural network (namely as ibUNet: https://github.com/harleyhk/ibUNet), which can accurately and efficiently evaluate the routability prediction in the following two tasks: the routing congestion prediction and the DRC hotspot prediction. Then, we made a comparison between our proposed model and the classic routability prediction model, RouteNet [2] on the newly released dataset CircuitNet [1] (https://github.com/circuitnet/CircuitNet), and found that our proposed model achieves promising rate reduction in terms of Avg-NRMSE, and consistently maintains better results on the metric of SSIM.

The rest of the paper is organized as follows. Section II reviews the related work. Section III presents the whole flow and details of our proposed model. Section IV provides measure definitions and metrics and conducts experiments to validate our methods. Section V gives conclusions for the work.

## II. RELATED WORK

### A. Prediction Modeling for EDA

In the past decade, using machine learning (ML) models to tackle the routability prediction tasks in the physical back-end design stage has been extensively investigated. Since the layout of a chip's physical design can be naturally viewed as an image, the existing classic convolutional neural network (CNN) models can be employed as the backbone of ML models for these tasks. These CNN-based models take image-like features in layouts as input feature maps and then take routing congestion maps [2, 3] or distribution maps of DRC hotspots [14] as the output labels. Some other researchers also tried to investigate the topological correlation among the circuit netlist by modelling the netlist as graph nodes and relation edges, where the graph neural networks (GNNs) models [15, 16] can be employed to learn the latent relations. Researchers also attempted to combine the advantages of both CNNs and GNNs to simultaneously capture the pixel-wise geometric position and topological relations [17].

Recently, the publicly available CircuitNet [1] dataset made a complement to the previous ISPD [25] contest benchmark suites. CircuitNet [1] overcame the obstacles in benchmarking and reproducing previous work in this field due to the lack of large public datasets. Like the ImageNet [6] dataset in computer vision research, CircuitNet [1] allows more research and verification to be conducted on a large-scale public dataset.

The two classic routability prediction tasks are defined in the following formula (1) and (2), in which a layout is divided into $w \times h$ tiles, and $f_{RC}$ and $f_{DRC}$ denote the channel sizes of the man-crafted input feature maps for RC and DRC, respectively.

**Routing Congestion Prediction**: A model is defined as a given layout design to predict the output routing congestion map based on the overflow levels after global routing.

$$m_{RC}: X^{RC} \in \mathbb{R}^{w \times h \times f_{RC}} \to Y \in \mathbb{R}^{w \times h} \quad (1)$$

**DRC Hotspot Prediction**: A model is built for a given routed design to predict the map of DRC hotspots, which are the locations of all DRC hotspots in a placement.

$$m_{DRC}: X^{DRC} \in \mathbb{R}^{w \times h \times f_{DRC}} \to Y_i \in \{0, 1\}^{w \times h} \quad (2)$$

These two models share similar architecture (with slight differences in input channel sizes). The model $m_{RC}$ takes input features map $X^{RC}$ (Macro Region, RUDY and Pin RUDY) and outputs a map with real values. The model $m_{DRC}$ takes input features map $X^{DRC}$ (Macro Region, RUDY, Pin RUDY, Cell density, and Congestion) and outputs a binary map.

### B. Prior Classic Model

RouteNet [2] is the first CNN-based model to evaluate the overall routability and predict the DRC hotspot map of a chip placement solution. The CNN-based models can be lightweight and are convenient for trying different operators and parameter recipes to adapt to the data distribution. Also, since model design and feature engineering are separate, the CNN-based models can be easily fine-tuned iteratively.

### C. Public Dataset for EDA

In recent years, with the development of artificial intelligence (AI) in the EDA field, various ML models have been invented to enhance the performance of traditional EDA procedures. However, the lack of large-scale public datasets hinders the reproducing of previous work and the conducting of benchmark comparisons. CircuitNet [1] is the first published large-scale open-source dataset for AI applications in this field. We conducted experiments on CircuitNet [1], and compared our proposed model with the classic model RouteNet [2].

## III. INCEPTION BOOSTED U-NET MODEL

### A. Routability Prediction

To evaluate routability in terms of routing congestion or DRC hotspots, a layout (or placement solution) is subdivided into an array of grid cells, each with the size of $l \times l$. Then, a rectangular layout with the size of $W \times H$ is divided into $w \times h$ grid cells, where $w = W/l$ and $h = H/l$.

We attempt to solve these two problems: routing congestion prediction and DRC hotspot map prediction. Both routing congestion and DRC hotspots are calculated at a grid-level granularity, like the image segmentation problem at pixel-level. Therefore, based on the analysis, proposed methods can utilize image-like features to train models, such as fully convolutional network (FCN) and U-Net (see our proposed model in Fig. 1), thereby effectively converting prediction task into an image-to-image segmentation or translation task.

### B. Main Architectures

Computer vision is an exciting application of artificial intelligence, and the two most common computer vision tasks are image classification [7, 8, 9, 10] and object detection [22, 23]. Fully convolutional networks (FCN) [20], where there is no fully connected (FC) layer at the end of the model, were first proposed to perform end-to-end semantic segmentation [20], outputting images with the same size of the original input.

In order to compensate for the impact of the down-sampling operator, a transposed convolution operator (ConvTranspose) or an up-sampling operator (e.g., nearest neighbor, bilinear, or bicubic) is usually added to get an up-sampled feature map and to control the final output size to be equal to the original size. This architecture is widely used in computer vision tasks. In such an encoder-decoder architecture, down-sampling is done by the encoder, and up-sampling is done by the decoder. U-Net [21] is a variant of FCN that also shares the same encoder-decoder architecture. To compensate for the information loss in the down-sampling operation, the skip connection of ResNet [9] is adopted, which combines feature maps in the encoder and decoder to help restore fine-grained details in the decoder in those FCN and U-Net models. U-Net is an encoder-decoder convolutional neural network with extensive applications in computer vision. Particularly in the image segmentation field, U-Net achieves huge improvements over previous state-of-the-art methods. U-Net consists of contraction paths and expansion paths. The contraction paths contain the architecture of convolution followed by a max-pooling operation with stride 2 for down-sampling. The expansion path consists of an up-sampling operator followed by convolution. The vanilla U-Net network [21] contains a total of 23 convolutional layers.

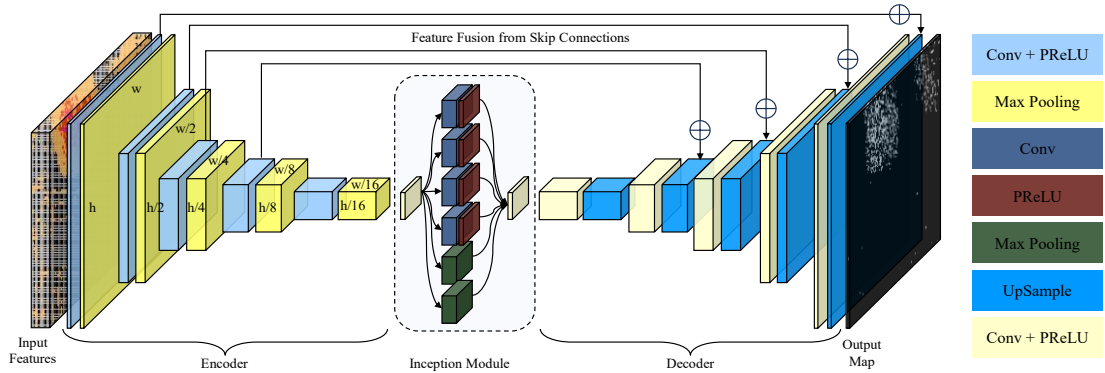

Fig. 1 The proposed Inception Boosted U-Net (ibUNet) model for routing congestion and DRC hotspot prediction tasks.

Due to their inherent architecture-based benefits, U-Net based neural networks show significant advantages in various image segmentation and other computer vision tasks. Since both routing congestion and DRC hotspot prediction are similar to pixel-wise image segmentation tasks, we intuitively use the latest state-of-the-art models in image segmentation to tackle these two problems. We tried to use vanilla U-Net [21] directly at the beginning but then abandoned it since it has too many parameters that leads to slow speed in training, besides, the training sample size is relatively small that causes over-fitting in the early stages of training. We analyzed the overfitting problem and believed that the feature dimension extracted by the current task is not high enough, thus, the vanilla U-Net and the later improved dense U-Net, such as U-Net++ [26], are unsuitable for solving the problem of either RC or DRC. If the feature dimension increases and the dataset size enlarges, more improved versions of U-Net can then be considered for use, especially for offline scenarios.

*C. The Improvements*

RouteNet [2] and Gpdl [3] share the same architecture and are typical CNN-based EDA customized deep learning models, which separate feature extraction from model design. The outstanding advantage of small models is that they can be quickly iteratively optimized in model design and feature extraction. Therefore, improving these models is a feasible research area that meets the industry's needs. Based on the original RouteNet, the proposed ibUNet model tries to use more layers of U-Net-like architecture but maintain the same level of parameters as RouteNet to achieve better performance. As shown in the workflow illustrated in Fig. 1, our proposed model ibUNet has 4 down-up sampling scales, in contrast, RouteNet contains 3. Based on the input feature map ($256 \times 256$), we designed 4 scales, which made the final down-scale size to be $16 \times 16$. The feature fusion from skip connections is kept as the main improvement from the ResNet [9] architecture, through which better performance can be achieved. Furthermore, an Inception module is introduced in the middle, acting as a zoom-in filter at the bottleneck part of network; this novel modification boosts the performance with a few extra operations.

Meanwhile, we conducted ablation experiments to optimize the parameters of the model since some operators are sensitive to data distribution: (1) We replaced the transposed convolution operator (ConvTranspose) with a bilinear interpolation for up-sampling to achieve the same effect with less computation; (2) We replaced ReLU and LeakyReLU with PReLU in the whole model as the activation function, that can achieve improvement since PReLU can adopt different data distributions of different channels in the tensors; (3) We found that due to different distributions of the input feature map and labelled data in these two tasks, batch normalization (BatchNorm) is more suitable for routing congestion prediction, while instance normalization (InstanceNorm) is more satisfactory for DRC hotspot prediction.

*D. The Novelty*

Although circuit design is complex, it can be analyzed by examining whether there are possible topology relationships between two nodes. In FCN networks, such as RouteNet [2], small convolution kernels (3x3) cannot explore distant topology relationships. Therefore, we inserted the Inception module (see Fig. 2) in the middle of U-Net [21], which is the bottleneck position of the network (see Fig. 1), to expand the receptive field to capture global relationship among netlist nodes while keeping the model lightweight.

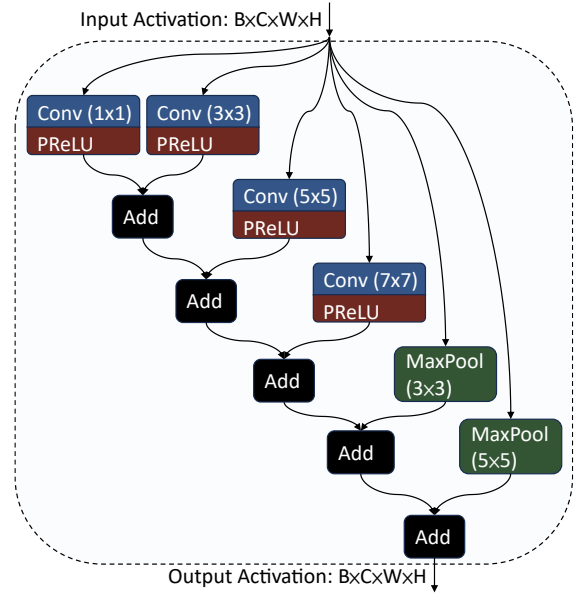

Fig. 2 The designed vanilla Inception module contains six filters and the feature maps are fused with element-wise Add operators. (B: batch, C: channel, W: width, H: height)

Stacking large convolution operations is computationally expensive in some prior classic CNN architectures, such as AlexNet [7] and VGGNet [8]. Choosing the correct kernel size for convolution operations becomes difficult as there are large variations in information location. Normally, larger kernels can obtain more global features over the image; on the other hand, smaller kernels are able to detect local features. Since extracting variable-sized features by determining a fixed kernel size is difficult, we need to design kernels of different sizes.

Instead of simply going deeper in terms of the number of layers, GoogLeNet [10] goes wider by proposing a parallel architecture named the Inception module, in which multiple kernels of different sizes are implemented within the same layer. Although there are some upgraded versions, such as Inception-V2 [11], Inception-V3 [12] and Inception-V4 [13], we designed a vanilla Inception module based on the original version from GoogLeNet [10] with following recipe filters: Conv ($1\times1$), Conv ($3\times3$), Conv ($5\times5$), Conv ($7\times7$), MaxPool ($3\times3$) and MaxPool ($5\times5$) filters. Most inception module-based networks use the Concatenate operator to fuse all outputs from the module's filters. After comparative experiments, we chose to fuse the output features with Elementwise-Add operators because we found that the performance of using Concatenate is the same as Elementwise-Add. Since using Concatenate operator requires constructing additional layers to convert the expanded channel size back to the original size, from this point of view, the Elementwise-Add solution can obtain less computation while keeping the non-linear property to benefit the generalization of the model. The time-consuming filters, such as Conv (5×5) and Conv (7×7) lead to heavy computation. To balance performance and model size, we inserted the Inception module into the neck position where the feature map is down-sampled 16 times (e.g., from $256 \times 256$ scales to $16 \times 16$).

IV. EXPERIMENTS

Since the public large-scale database CircuitNet dataset was released only recently, the comparative evaluation of algorithms in the field of chip placement and routing had been hindered due to the unavailability of open-source code and public datasets. In this paper, we present our proposed algorithm, ibUNet, and will compare it with RouteNet [2], which offers open-source code (https://circuitnet.github.io) and is the only work that we can find that is tested in the CircuitNet dataset.

*A. Task and Dataset*

Our study focuses on two major prediction tasks: Routing Congestion (RC) prediction and Design Rule Checking (DRC) hotspot map prediction. RC occurs when the routing demand exceeds available routing resources, while DRC refers to the location violation of design rules during the routing phase. To facilitate our evaluation, we utilized the public CircuitNet [1] dataset, which collected 3 variations of datasets, CircuitNet-N28, CircuitNet-N14 and ISPD2015 from runs of commercial design tools and also provided baselines for RC prediction (RouteNet [2]) and DRC hotspot prediction (Gpdl [3]). RouteNet and Gpdl use similar fully CNN-based EDA customized deep learning models that combine the RUDY map, pin RUDY map, and the macro-region as input features for RC prediction. Meanwhile, RouteNet is a customized network for DRC hotspot prediction. In addition to the three hand-crafted features mentioned above, it also incorporates cell density and routing congestion map as input. We evaluated our proposed model on the CircuitNet dataset for these two cross-stage prediction tasks: RC prediction and DRC prediction. In this paper, all the experiments were conducted in the CircuitNet-N28 (training samples: 7078, and testing: 3164), which is based on RISC-V designs and 28nm planar technology and provides most complete support for both RC and DRC hotspot predictions.

*B. Features for Task*

**Routing Congestion Features:** the input features are Macro Region, RUDY, and Pin RUDY, and the dimension size is: $256 \times 256 \times 3$.

**DRC Hotspot Features:** the input features are Macro region, RUDY, Pin RUDY, Cell density, and Congestion map; the dimension size is: $256 \times 256 \times 9$.

**Feature Definitions:**

(1) **Macro regions** are the regions covered by macros on a layout (layout solution) that are fixed and occupied routing resources, used to estimate the relative distribution of routing resources available in each tile.

(2) **RUDY** (Rectangular Uniform wire DensitY) [24] is a measure of each net node over spatial dimension and can be treated as an early estimate of routing congestion after placement. RUDY is widely used for its high efficiency and accuracy.

(3) **Pin RUDY** is a pin variant named Pin RUDY from [2] and [3], calculated based on each pin and the net node to which the pin is connected, and it is an estimation of the pin density.

(4) **Cell density** is the density distribution of cells, equal to the cell counts in each tile.

(5) **Congestion map** is the overflow or utilization of routing resources in each tile, which is computed by the routing resources. The report contains three parts: total tracks, remaining tracks, and overflow based on each tile. Routing wires must be routed on tracks; thus, tracks are equivalent to routing resources.

*C. Metrics for Task*

**Metrics 1**: Normalized Root-Mean-Square-Error (NRMSE).

In statistical modeling, particularly regression analyses and computer vision tasks like image restoration [27], Root-Mean-Square-Error (RMSE) [18] is a common way to measure the quality of a model fitting by estimating the pixel-wise difference between label data $y$ and predicted value $\hat{y}$. NRMSE (as defined in Eqn. (3)) furthers the results of root finding and normalization to compare samples of different scales and distributions. There are several normalization methods, here we use the min-max normalization. The smaller the NRMSE value, the better the model performance.

$$NRMSE = \frac{1}{y_{max}-y_{min}} \sqrt{\frac{\sum_{i=1}^{h}\sum_{j=1}^{w}(y_{i,j}-\hat{y}_{i,j})^2}{h \times w}} \qquad (3)$$

**Metrics 2**: Structural Similarity Index Measure (SSIM).

Another metric exploited is the Structural Similarity Index Measure (SSIM) [1, 19]. Unlike RMSE [18] or PSNR [18] (Peak Signal-to-Noise Ratio), which reckons absolute errors, SSIM is a perceptual-based model that considers image degradation as perceptual changes in structural information, which is widely employed in image processing [27]. Structural information refers to the strong interdependence between pixels, especially when they are spatially close. The higher the SSIM value, the better the model's performance on fitting.

**Metrics 3**: Area Under the Curve (AUC).

The Area Under the Curve (AUC) [1, 2] is a measurement to assess the ability of a binary classifier and is used as a summary of the Receiver Operating Characteristic (ROC) [1, 2] curve. The higher the AUC value, the better the model's performance in distinguishing between the positive and negative classes. Due to the variable distinguishing thresholds, the ROC curve is a graphical representation that illustrates the diagnostic capability of the binary classifier system. In the meantime, it can be used to identify the optimal threshold for binary classification. According to the ROC curve method, after calculating the Confusion matrix [1, 2] values of different thresholds and drawing the ROC curve, we can determine the optimal threshold of the training model, i.e., the top-left point in the ROC curve.

*D. Comparative Experiments*

We evaluated the RouteNet [2] model and our proposed ibUNet model with two objective evaluation metrics: Avg NRMSE and Avg SSIM on the routing congestion prediction task. As all the scores shown in TABLE I, these experimental results prove that our proposed method can achieve up to 5% rate reduction in terms of Avg-NRMSE and consistently achieves better performance on the metric of SSIM compared with RouteNet.

TABLE-1: EXPERIMENTAL RESULT ON ROUTING CONGESTION PREDICTION

| Congestion | Avg NRMSE | Avg SSIM |
|---|---|---|
| RouteNet | 0.0493 | 0.7805 |
| ibUNet | 0.0467 | 0.7978 |

The DRC hotspot prediction task can be considered a detection or segmentation task in the field of computer vision since the output is an activation tensor map with the same size as the original input feature map. At the same time, each element in the output binary tensor map can be regarded as a binary classification. Since it is a binary classification problem, the false alarm (i.e., the non-hotspots are predicted as hotspots or the hotspots are predicted as non-hotspots by the classifier) problem needs to be seriously considered, so metrics such as the Confusion matrix [1, 2], ROC and AOU [1, 2] shall be involved.

TABLE-2: EXPERIMENTAL RESULT ON DRC HOTSPOT PREDICTION

| DRC | Avg NRMSE | Avg SSIM | AUC of ROC |
|---|---|---|---|
| RouteNet | 0.0382 | 0.9711 | 0.9391 |
| ibUNet | 0.0304 | 0.9762 | 0.9474 |

We evaluated the RouteNet [2] model and our proposed ibUNet model with three objective evaluation metrics: Avg NRMSE, Avg SSIM, and AUC of ROC on the DRC hotspot prediction task. As all the scores shown in TABLE II, these experimental results prove that our proposed method can achieve up to 20% rate reduction in Avg-NRMSE and consistently achieves better results on the metric of SSIM and AUC of ROC compared with RouteNet.

*E. Incremental Experiments*

To further explore models' performance, we plot all trained models' performance curves with 200 different training epochs (1 epoch =1000 iterations), as shown in Figure 3 to 6.

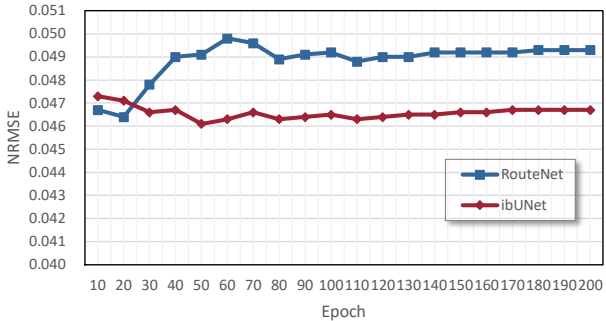

Fig. 3: NRMSE of the congestion prediction task.

As shown in Fig. 3 and Fig. 4, Our proposed model ibUNet can stably outperform the model RouteNet [2] after about 30 training epochs.

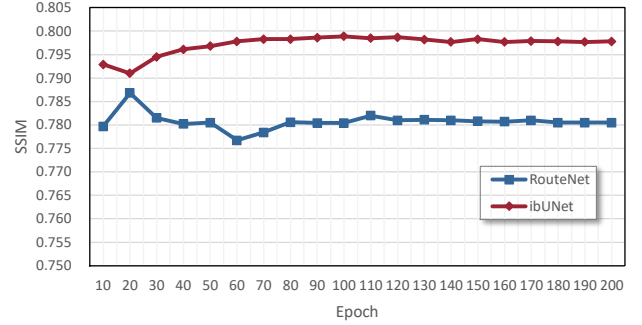

Fig. 4: SSIM of the congestion prediction task.

Fig. 3 and Fig. 4 also reveal that the model RouteNet [2] has overfitting on the routing congestion prediction dataset after a few epochs of training, and our proposed model ibUNet does not suffer from this problem.

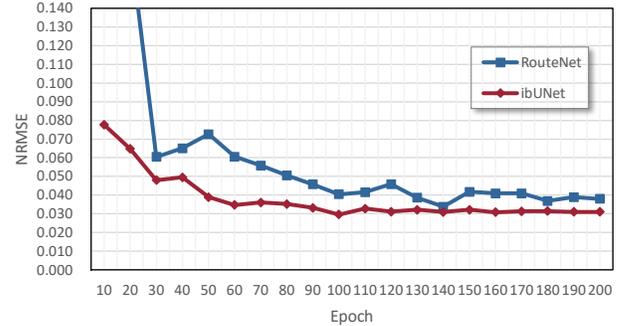

Fig. 5: NRMSE of the DRC hotspot prediction task.

Fig. 5 and Fig. 6 demonstrate that for all models trained in the DRC hotspot prediction task, our proposed model ibUNet consistently outperforms the model RouteNet [2], while the trained models also converge at a faster speed.

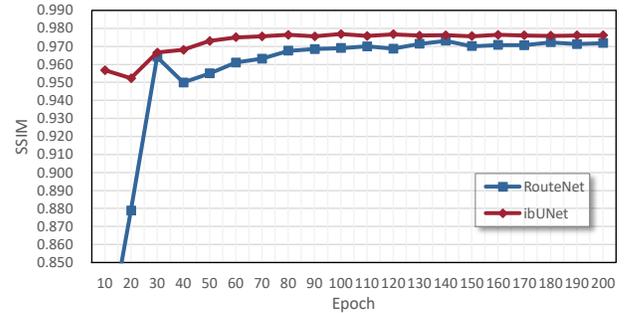

Fig. 6: SSIM of the DRC hotspot prediction task.

As shown in Fig. 5 and Fig. 6, neither model RouteNet [2] nor ibUNet are having overfitting problem in the DRC hotspot prediction dataset. All these four figures show that the ibUNet model can achieve stable performance and relatively faster fitting after about 40 training stages.

In all experiments, we set the batch size to be 16 for both models. To get reliable and fair models for comparison, both the RouteNet [2] and our ibUNet models are trained in an AMD

64bit desktop machine with CPU: AMD Ryzen 9 3900X 12-Core Processor + 64G memory, and one GPU: NVIDIA GeForce GTX 3070 + 8G GPU memory.

## V. Conclusions

We propose a novel U-Net architecture with an Inception module inside for routability prediction tasks in the chip design flow to prevent routing congestion and DRC violations in the early stages. It can achieve promising enhancement compared with the prior classic model.

However, it should be noted that the separation of feature engineering and model design limits our current approach. The current man-crafted image pixel-like features make it difficult to encode topological relationships among netlist nodes. Combining the GNN-based and CNN-based methods and designing new features to explore the topological relationships among netlist nodes are worth future research. Also, designing a new hybrid loss function and other neural network techniques for the training process may enhance the model's performance.

Last but not least, the proposed model can be used for other image semantic segmentation-based tasks, such as medical image diagnosis and self-driving cars utilizing lightweight edge computing devices.